\documentclass[]{emulateapj}

\def\rmit#1{{\it #1}}              

\def\eg{\rmit{e.g.}}

\usepackage{natbib,graphicx}

\usepackage{color}

\shorttitle{Local helioseismology of sunspot models}

\shortauthors{Felipe et al.}

\begin{document}

\title{Evaluation of the capability of local helioseismology to discern between monolithic and spaghetti sunspot models}

\author{T. Felipe\altaffilmark{1}, A. D. Crouch\altaffilmark{1}, A. C. Birch\altaffilmark{2}}
\email{tobias@nwra.com}

\altaffiltext{1}{NorthWest Research Associates, Colorado Research Associates, Boulder, CO 80301, USA}
\altaffiltext{2}{Max-Planck-Institut f\"{u}r Sonnensystemforschung, Justus-von-Liebig-Weg 3, 37077 G\"{o}ttingen, Germany}

\begin{abstract}
The helioseismic properties of the wave scattering generated by monolithic and spaghetti sunspots are analyzed by means of numerical simulations. In these computations, an incident $f$ or $p_1$ mode travels through the sunspot model, which produces absorption and phase shift of the waves. The scattering is studied by inspecting the wavefield, computing travel-time shifts, and performing Fourier-Hankel analysis. The comparison between the results obtained for both sunspot models reveals that the differences in the absorption coefficient can be detected above noise level. The spaghetti model produces an steep increase of the phase shift with the degree of the mode at short wavelengths, while mode-mixing is more efficient for the monolithic model. These results provide a clue for what to look for in solar observations to discern the constitution of sunspots between the proposed monolithic and spaghetti models.

\end{abstract}



\section{Introduction}

The internal structure of sunspots has been an enduring question in solar physics. Two competing scenarios have been proposed several decades ago, the so-called ``monolithic sunspot'' and the ``spaghetti sunspot'', but still there is no definitive evidence which can end the debate and prove one of the models wrong. The monolithic model \citep{Cowling1953} considers the sunspot magnetic field as approximately homogeneous below the visible solar surface. Due to some difficulties of this model to explain several solar phenomena, for example umbral dots, \citet{Parker1979} proposed an alternative model assuming that the sunspot is composed of a group of many magnetic flux tubes bundled together, the so-called spaghetti model. This model naturally explains umbral dots as the signature of the field-free hot gas surrounding the flux tubes. However, several works \citep[\eg,][]{Weiss+etal1990,Hurlburt+etal1996,Weiss2002,Schussler+Vogler2006} have claimed that umbral dots can be included in the monolithic model as the result of magnetoconvection. The analysis of umbral dots through direct observations of the solar photosphere using high-resolution Stokes polarimetry has not been able to discern between these two models \citep{Lites+etal1991, SocasNavarro+etal2004}. 

Recently, there has been a substantial development in realistic three-dimensional magnetohydrodynamic numerical simulations of sunspots, including radiative transfer and a realistic equation of state \citep{Rempel+etal2009, Rempel2011}. These simulations show over time a significant fragmentation of the subsurface magnetic field even if they are started from a monolithic initial state. Most major fragmentations become visible at the photosphere. \citet{Rempel2011} suggests that sunspots without light bridges are more monolithic than those which present light bridges and signs of flux separation.

Indirect observations of the solar interior by means of the interpretation of the acoustic wave field in and around sunspots seem to be the most promising approach to unveil the subsurface structure of sunspots. The papers by \citet{Braun+etal1988, Braun+etal1992} have proven the strong interaction of the $p-$modes with solar magnetic fields. They found that sunspots can absorb up to half of the incident $p-$modes power. The absorption and scattering phase shift produced by sunspots show a dependence with the frequency, degree, radial order, and azimuthal order \citep{Bogdan+etal1993, Braun1995}. In order to use this information to interpret the properties of magnetic structures, it is necessary to develop theoretical models for the interaction of waves with magnetic fields. Ideally, these models will reveal some differences in the way that acoustic waves interact with monolithic and spaghetti models, indicating which solar observations can distinguish between the two scenarios.

Mode conversion \citep{Cally+Bogdan1993} is the most promising mechanism to explain the observed absorption in sunspots. It occurs because in magnetized regions the slow and fast wave modes are coupled and can exchange energy, especially around the height where the sound and Alfv\'en speeds are similar. Below this height the sound speed greatly exceeds the Alfv\'en speed, and the fast, slow, and Alfv\'en waves are decoupled. In these deep layers the fast wave is similar to an acoustic wave in non-magnetic regions, while the slow and Alfv\'en modes are incompressive waves which travel along field lines. According to the mode conversion hypothesis, absorption is produced by the conversion of incident $p-$modes into these waves in magnetized regions, which propagate downward extracting energy from the acoustic cavity. \citet{Cally+etal1994} showed that mode conversion produced by vertical magnetic fields can account for the absorption of $f-$modes, but in order to explain the absorption of $p-$modes it is necessary to include inclined magnetic fields \citep{Crouch+Cally2003, Cally+etal2003}.     

In the case of thin flux tubes, it is well established that they can support sausage and kink modes \citep{Roberts+Webb1978, Spruit1981}. These tube waves are excited by the interaction of the flux tube with the incident acoustic waves, and they travel upward or downward removing energy from the $p-$modes \citep{Bogdan+etal1996, Crouch+Cally1999, Hindman+Jain2008}. Several works \citep{Hanasoge+etal2008, Hindman+Jain2012, Felipe+etal2012a} evaluated the absorption produced by isolated thin flux tubes. However, following the hypothesis of the spaghetti model, sunspots are composed by bundles of these elements. \citet{Bogdan+Fox1991} realized the relevance of multiple scattering to account for the interaction of sound waves with groups of flux tubes, and analyzed pairs of flux tubes in unstratifed atmospheres. \citet{Keppens+etal1994} found that spaghetti sunspots are much more efficient absorbers than monolithic models, and the study of \citet{Hanasoge+Cally2009} of pairs of flux tubes in gravitationally stratifed atmospheres supports their results. Recently, \citet{Felipe+etal2013} evaluated the scattering of the $f$ mode produced by groups of flux tubes, and found that multiple scattering tends to increase the absorption and reduce the phase shift, while \citet{Daiffallah2014} obtained that a loose cluster of tubes is a more efficient absorber than a compact cluster or an equivalent monolithic tube.

As of today, no study has analyzed the scattering produced by a sufficient number of tubes to construct a model comparable to actual sunspots. However, this is now feasible thanks to numerical simulations. The objective of this paper is to compare the seismic response produced by spaghetti models with that obtained for monolithic sunspots. As discussed previously, the mechanisms that produce the absorption in these two cases may be different. On the one hand, for the thick flux tube representative of monolithic sunspots, fast and slow magnetoacoustic waves are strongly coupled around the region where the sound and Alfv\'en speed are similar, and slow magnetic modes extract energy from the incident acoustic waves. On other hand, in the case of spaghetti sunspots, the absorption is produced by the excitation of tube waves in the constituent flux tubes and enhanced by multiple scattering. In this work, we do not attempt to identify the wave modes that generate the absorption \citep[see][]{Felipe+etal2012a}. We aim to find some differences in the measurements obtained from these two distinct magnetic field configurations, which could be used as a constraint to discern between these models.

The layout of this paper is as follows. In Section \ref{sect:numerical} we present the numerical simulations and the sunspot models. Section \ref{sect:wavefield} describes the wavefields produced by both sunspots. In the next two sections we discuss the results of the helioseismic measurements performed, including the travel-time shift (Section \ref{sect:traveltimes}) and Fourier-Hankel analysis (Section \ref{sect:hankel}), while in Section \ref{sect:mode_mixing} we evaluate the mode-mixing. Finally, the conclusions of these results are found in Section \ref{sect:conclusions}.

\section{Numerical simulations}
\label{sect:numerical}

\begin{figure}[!ht] 
 \centering
 \includegraphics[width=9cm]{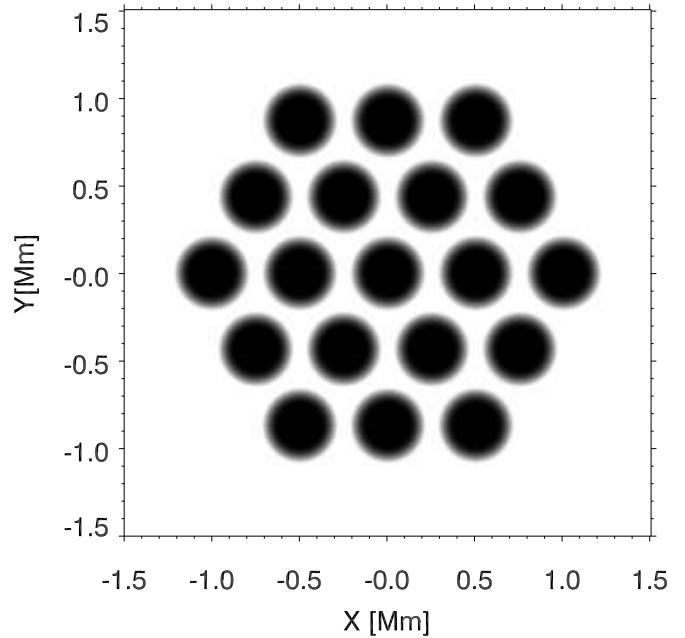}
  \caption{Distribution of the 19 inner flux tubes in the spaghetti sunspot model.}
  \label{fig:tubos_pos}
\end{figure}

MHD simulations were carried out using the Mancha code \citep{Khomenko+Collados2006, Felipe+etal2010a}. We used two different magnetohydrostatic backgrounds. The monolithic sunspot was constructed following \citet{Khomenko+Collados2008}. The photospheric magnetic field at the axis of the spot is 1500 G. It smoothly decreases with radius, and at 12 Mm from the center of the sunspot the magnetic field vanishes and the atmosphere corresponds to CSM\_B quiet Sun model \citep{Schunker+etal2011}. The spaghetti sunspot is composed by a bundle of 1663 identical flux tubes inside a circle with radius of 10.7 Mm, with a minimum distance between the center of the tubes of 0.5 Mm. Figure \ref{fig:tubos_pos} shows the distribution of the 19 tubes closest to the center of the sunspot as an example. The same pattern is repeated until filling the whole area of the sunspot. Each individual tube model is taken from \citet{Khomenko+etal2008a}. Within a radius of 0.1 Mm they have a photospheric magnetic field strength of 2500 G and small variations in the thermodynamic variables. In the next 0.1 Mm the magnetic field is reduced to zero. The atmosphere surrounding the tubes also corresponds to CSM\_B. The number of tubes was selected in order to match the photospheric magnetic flux of the monolithic model. 

\begin{figure}[!ht] 
 \centering
 \includegraphics[width=9cm]{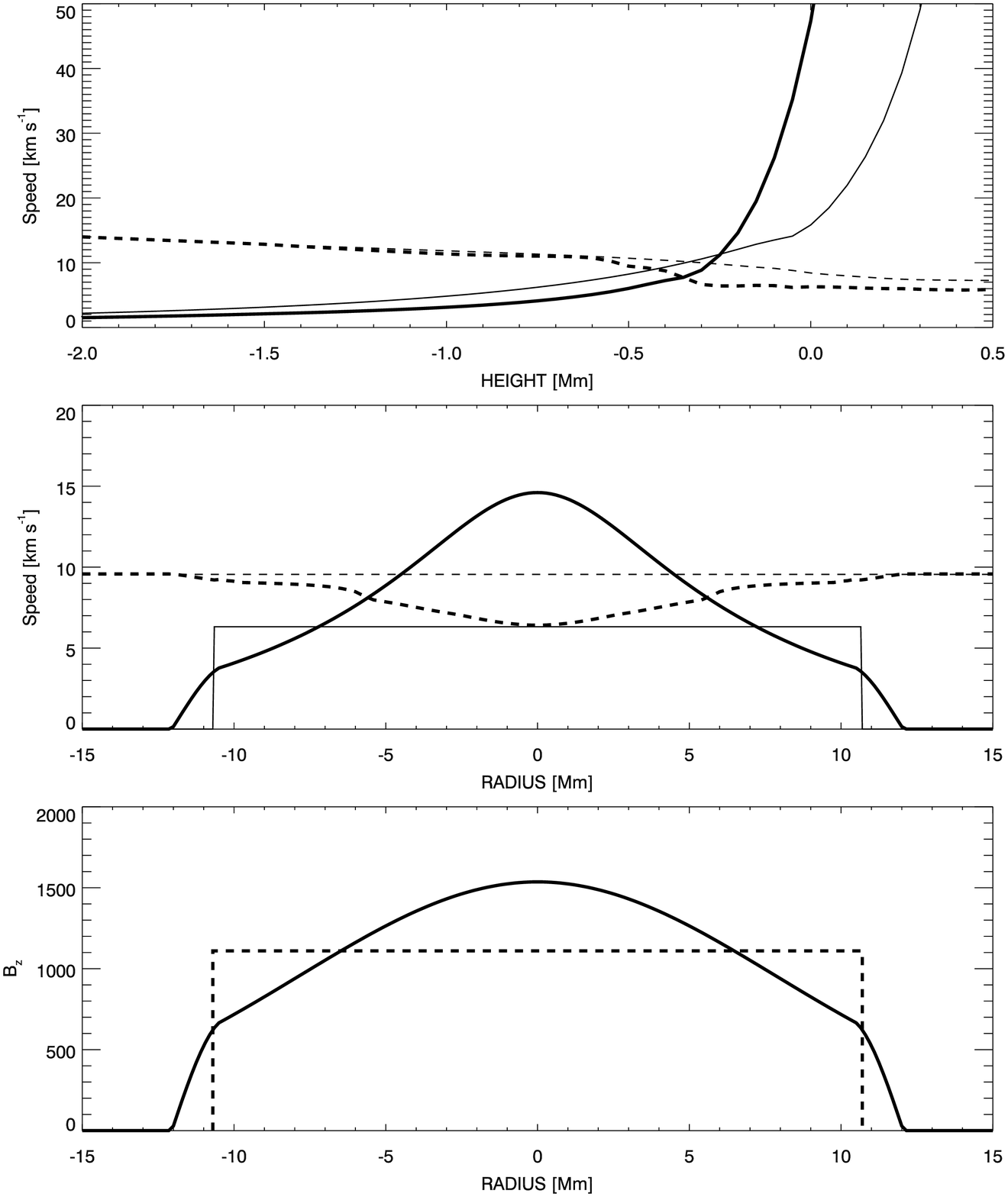}
  \caption{Top panel: Characteristic speeds at the center of the sunspot models as a function of height. Middle panel: Characteristic speeds at $z=-0.2$ Mm as a function of radius ($z=0$ is photosphere). Both panels show the Alfv\'en speed (solid lines) and sound speed (dashed lines) of the monolithic model (thick lines) and the spaghetti model (thin lines). In the case of the middle panel, the Alfv\'en speed of the spaghetti model corresponds to that obtained if the magnetic flux were uniformly distributed inside the radius of the sunspot. Bottom panel: Photospheric vertical magnetic field as a function of radius. The solid line corresponds to the monolithic sunspot, and dashed line to the magnetic field strength of the spaghetti sunspot if the magnetic flux were uniformly distributed inside the radius of the spot.}
  \label{fig:velocities_B_models}
\end{figure}

Figure \ref{fig:velocities_B_models} shows a comparison of the characteristic velocities and magnetic field between both sunspot models. The monolithic sunspot has a Wilson depression with a reduction of the sound speed in the inner region of the sunspot. The spaghetti model lacks a Wilson depression. All the flux tubes in the spaghetti model, from the axis of the sunspot to the outer radial distances, have the same variation in thermodynamic variables, and do not produce a significant modification of the sound speed. The top panel also shows that in the deeper layers the Alfv\'en speed at the axis of one of the flux tubes of the spaghetti model is higher than that on the axis of the monolithic sunspot. On the other hand, around the photosphere the Alfv\'en speed of the monolithic sunspot greatly exceeds that of the flux tubes due to the decrease of the density produced by the Wilson depression. The figure also reveals the differences in the distribution of the magnetic field between the sunspot models. In order to illustrate a meaningful comparison between the whole picture of both sunspots, in the case of the spaghetti model we have plotted a constant magnetic field inside the radius of the sunspot which generates the same magnetic flux of the model, instead of an alternation of small regions with high magnetic flux (flux tubes) and with no magnetic flux (magnetic-free regions between tubes). The Alfv\'en speed of the spaghetti sunspot from the middle panels was obtained using this uniform distribution of the magnetic field. In this work we neither aim to study realistic sunspot models nor reproduce observations. We present some simple sunspot models that account for some basic properties of monolithic and spaghetti sunspots with the same magnetic flux as a first step to evaluate the differences in the way that those structures interact with acoustic waves.

We have performed independent simulations for wave packets with fixed radial order, including $f$ and $p_1$ modes. Perturbations in velocity, pressure, and density were imposed as initial conditions to produce an $f$ or $p_1$ wave packet which propagates in the ${\bf +\hat{x}}$ direction \citep{Cameron+etal2008, Felipe+etal2012a}. The eigefunctions were computed using the method outlined in Section 3.1 of \citet{Crouch+etal2005}. Both wave packets were introduced in both sunspot models, so the analysis presented in this paper has been obtained from a set of four simulations. The horizontal computational domain is $x \in [-44.5,27.5]$ Mm and $y \in [-27.5,27.5]$ Mm, with the center of the sunspot model located at $x=y=0$. The horizontal spatial step for the spaghetti sunspot is 50 km, while the numerical simulations of the monolithic model has a three times coarser resolution. The top boundary is located at 1 Mm above the photosphere, and the depth of the bottom boundary depends on the radial order of the simulation. For the simulations of the $f$ mode it is located 6 Mm below the photosphere, and in the case of the $p_1$ mode it is at $z=-15$ Mm, with $z=0$ corresponding to the photospheric level. All simulations use the same vertical spatial step $\Delta z=50$ km. Equivalent two dimensional (2D) quiet Sun simulations were computed as a reference.

\section{Examination of the wavefield}
\label{sect:wavefield}

\begin{figure}[!ht] 
 \centering
 \includegraphics[width=9cm]{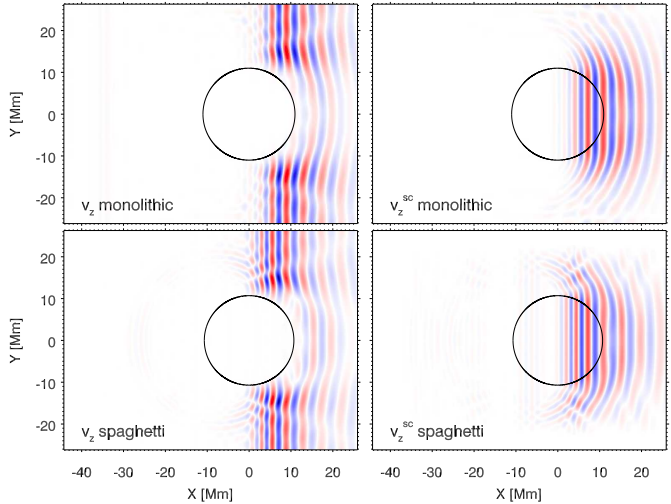}
  \caption{Photospheric vertical velocity of the $f$ mode after 105 min of simulation. Top panels: monolithic sunspot, bottom panels: spaghetti sunspot, left panels: full wavefield, right panels: scattered wave. The blue color indicates upwards flows and the red color shows downward flows. The black circle shows the outer limit of the sunspot model. }
  \label{fig:scattering_f}
\end{figure}

Figure \ref{fig:scattering_f} shows the vertical velocity at $z=0.2$ Mm produced 105 min after the start of the $f$ mode simulations. At this time step, the $f$ mode wave packet is almost finishing its travel across the sunspot. Both the monolithic and spaghetti sunspots illustrate some of the well known properties of the interaction between waves and solar magnetic structures \citep{Braun+etal1987,Braun1995, Cameron+etal2008}: firstly, the region of the wavefront which passes through the sunspot is shifted forward in space, since the waves travel faster in the magnetized atmosphere; secondly, the amplitude of those waves is significantly reduced inside the sunspot, and it remains smaller just behind the spot. The scattering wave $v_z^{sc}$ is obtained as the difference between the simulation with the sunspot being present and the equivalent quiet Sun simulation. The scattered wave field mainly propagates in the forward direction of the incident wave, although its wavefront is curved. From a visual inspection of the scattered wave produced by the monololithic and the spaghetti models some differences arise. In the case of the latter, most of the wavefront behind the bulk of the sunspot is in phase at constant $y$, while for the monolithic sunspot the curvature is more gradual.

\begin{figure}[!ht] 
 \centering
 \includegraphics[width=9cm]{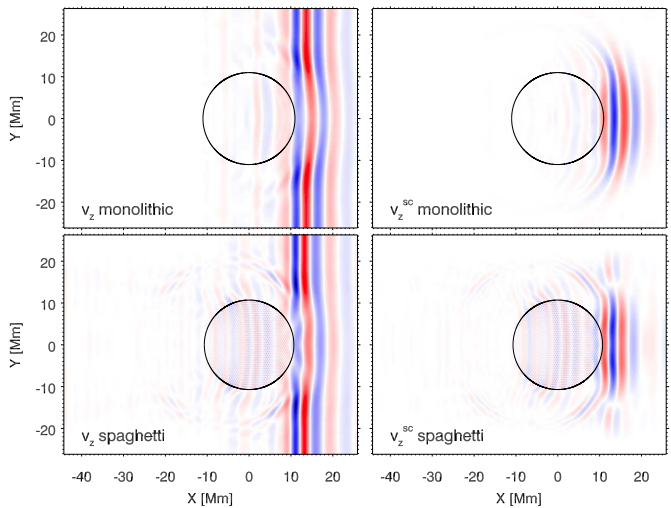}
  \caption{Vertical velocity of the $p_1$ mode after 80 min of simulation. Top panels: monolithic sunspot, bottom panels: spaghetti sunspot, left panels: full wavefield, right panels: scattered wave. The blue color indicates upwards flows and the red color shows downward flows. The black circle shows the outer limit of the sunspot model. }
  \label{fig:scattering_p1}
\end{figure}

Figure \ref{fig:scattering_p1} shows the full and scattered wavefield produced by the propagation of a $p_1$ mode through the sunspot models. Similar features can be seen. In this case the phase shift produced by the sunspots is smaller. For the $p_1$ simulations the power peak of the initial wave packet is located at lower $L$ (higher wavelength), and for those waves the phase shift is generally smaller \citep{Braun1995, Felipe+etal2012a}. A detailed comparison of the phase shift produced for $f$ and $p_1$ modes by monolithic and spaghetti sunspots will be presented in the following sections. For the $p_1$ mode simulations it is also noticeable the flat wavefront of the scattered wave produced by the spaghetti sunspot, in contrast to the curved scattered wave produced by the monolithic model. Another difference is the low amplitude radially outgoing scattered wave which appears in the spaghetti model wavefield. It is centered to the left of the axis of the sunspot, and it is generated after the first contact of the wave packet with the sunspot model.

\section{Wave travel-times}
\label{sect:traveltimes}

We define the travel-time in an analogous way to \citet{Gizon+Birch2002}. At each spatial point, the traveltime shift $\delta\tau (x,y)$ is calculated as the time lag that minimizes the difference between the photospheric vertical velocity in the sunspot simulation $v_z(x,y,t)$ and the quiet Sun reference vertical velocity  $v_z^{\rm qs}(x,t)$. This is equivalent to obtaining $\delta\tau$ as the time $\tau$ that maximizes the function

\begin{equation}
\label{eq:cross_correlation}
F(x,y,\tau)=\int v_z(x,y,t)v_z^{\rm qs}(x,t-\tau ){\rm d}t
\end{equation}

Note that quiet Sun data are obtained from 2D simulations, and the same reference wavefield is applied to every $y$ position. Observational works usually compute the travel time using the cross-correlation between the signal measured at two points of the solar surface. In our simulations this step is simplified and we use directly the velocity, since the initial wave packets in the sunspot simulations and the reference quiet Sun are the same. The wavevectors are parallel to the $x$ axis and all wavenumbers are in phase at $t=0$.  

We applied a filter to the vertical velocity wavefield in order to analyze some selected $L$ and frequency bands. We use two different filters, one applied to the $f$ mode simulations and the other to the $p_1$ mode simulations. The $f$ mode filter is centered at $L=1166$ and covers the range in $L$ values between 1020 and 1311, while the range in frequencies varies with $L$. It has a width of 0.7 mHz, with the center located at the frequency of the $f$ mode at the corresponding $L$ value. The $p_1$ filter is centered at $L=874$ and spans from $L=728$ to $L=1020$, with the same width in temporal frequency of the $f$ mode filter. We chose filters with similar frequency, so the $p_1$ filter includes lower $L$ values. The exact central $L$ was selected to allow a direct comparison with the results of the following section.   

\begin{figure}[!ht] 
 \centering
 \includegraphics[width=9cm]{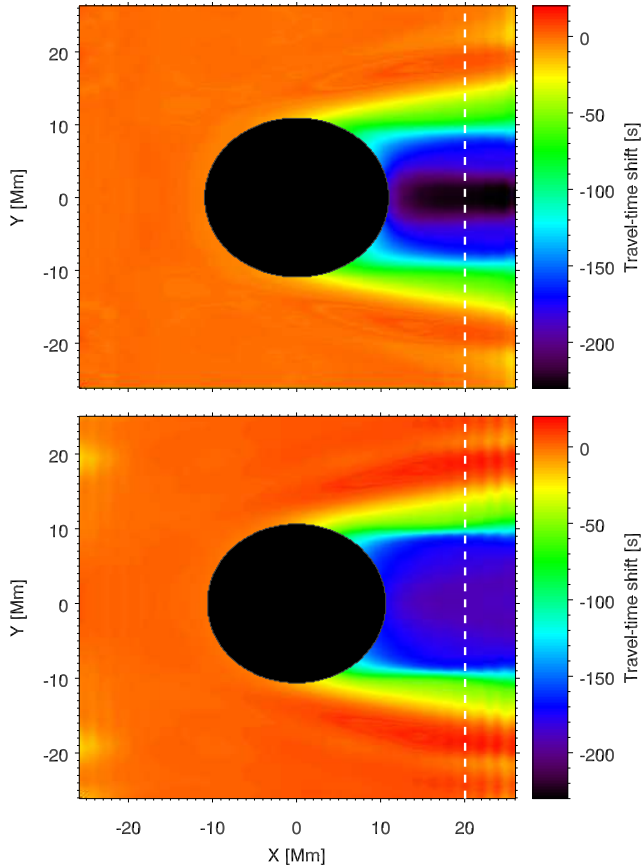}
  \caption{Travel-time shift of the $f$ mode centered at $L=1166$ (spanning from $L=1020$ to $L=1311$) for the monolithic sunspot (top panel) and the spaghetti sunspot (bottom panel). The black circles correspond to the area of the sunspot models. Vertical dashed white lines correspond to the position plotted in Figure \ref{fig:traveltime_plot_f}.}
  \label{fig:traveltime_map_f}
\end{figure}

\begin{figure}[!ht] 
 \centering
 \includegraphics[width=9cm]{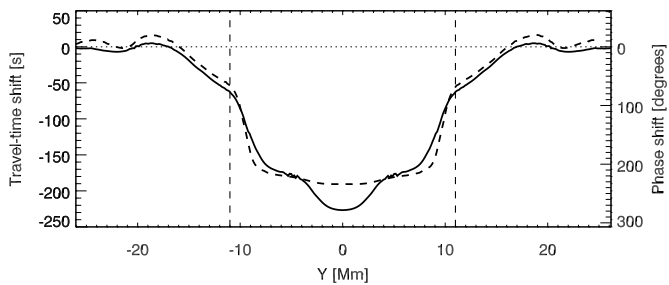}
  \caption{Travel-time shift of the $f$ mode centered at $L=1166$ (spanning from $L=1020$ to $L=1311$) as a function of $y$ at $x=20$ Mm. Solid lines represent the monolithic sunspot and dashed lines correspond the to spaghetti sunspot. Vertical dashed lines indicate the limits of the sunspot at $x=0$.}
  \label{fig:traveltime_plot_f}
\end{figure}

Figure \ref{fig:traveltime_map_f} shows spatial maps of the travel-time shifts obtained for the $f$ mode in the monolithic and spaghetti sunspots. In both cases negative travel-time shifts are measured behind the sunspot for all $y$ positions where the spot is present. For a better comparison, Figure \ref{fig:traveltime_plot_f} shows the variation of the travel-time with $y$ at $x=20$ Mm. The magnitude of the travel-time shifts produced by the spaghetti model behind the sunspot are lower than those produced by the monolithic model, which means that the waves which travel through the latter propagate more rapidly. For $y$ in the range between -9 and 9 Mm, the travel-time shift produced by the spaghetti model is almost constant, which agrees with the perception of flat scattered wavefront from Figure \ref{fig:scattering_f}. At further distances the travel-time shift is abruptly reduced. Some small positive travel-time shifts appear at around $|y|=19$ Mm, especially for the spaghetti sunspot. They are produced by the curvature of the scattered wave. Its interference with the incident wave can be seen in the left panels of Figure \ref{fig:scattering_f}, producing a shift of the wavefront towards the left.

\begin{figure}[!ht] 
 \centering
 \includegraphics[width=9cm]{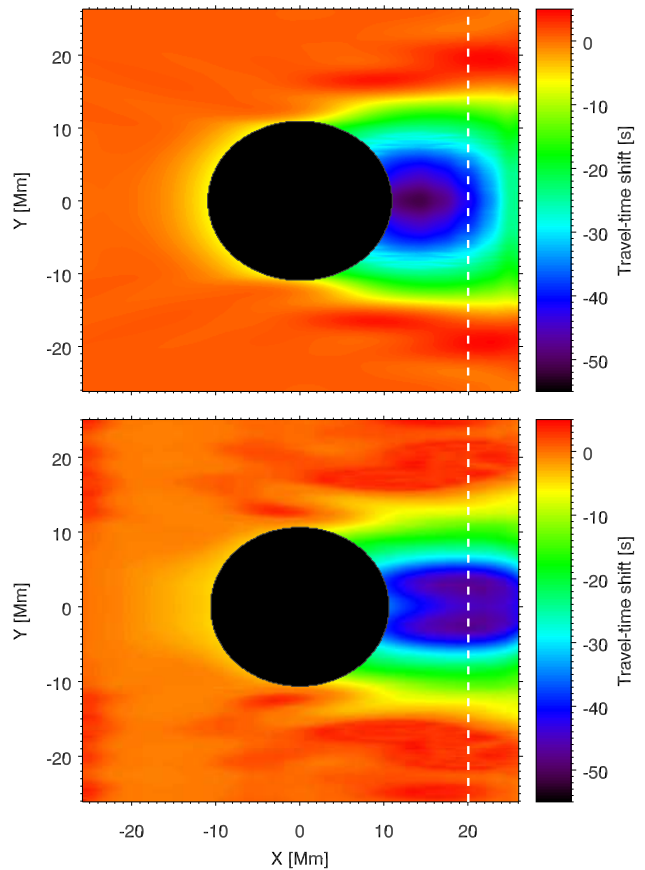}
  \caption{Travel-time shift of the $p_1$ mode centered at $L=874$ (spanning from $L=728$ to $L=1020$) for the monolithic sunspot (top panel) and the spaghetti sunspot (bottom panel). The black circles correspond to the area of the sunspot models. Vertical dashed white lines correspond to the position plotted in Figure \ref{fig:traveltime_plot_p1}.}
  \label{fig:traveltime_map_p1}
\end{figure}

\begin{figure}[!ht] 
 \centering
 \includegraphics[width=9cm]{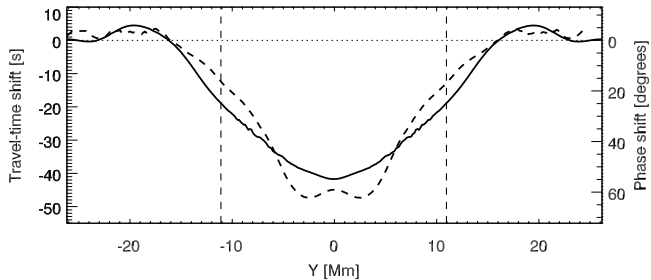}
  \caption{Travel-time shift of the $p_1$ mode centered at $L=874$ (spanning from $L=728$ to $L=1020$) as a function of $y$ at $x=20$ Mm. Solid lines represent the monolithic sunspot and dashed lines correspond to the spaghetti sunspot. Vertical dashed lines indicate the limits of the sunspot at $x=0$.}
  \label{fig:traveltime_plot_p1}
\end{figure}

The travel-time shift for $p_1$ mode shows some differences with the $f$ mode. Figures \ref{fig:traveltime_map_p1} and \ref{fig:traveltime_plot_p1} show that the travel-time shift is smaller. However, the different $L$ values used for the filtering must be taken into account. A detailed comparison of the phase shift will be shown in Section \ref{sect:phase}. Regarding the spatial distribution, the spaghetti model produces two lobes along the $y$ axis, with a less negative travel-time shift at $y=0$. The magnitude of the travel-time shift peaks behind the monolithic model and then decreases as $x$ increases. This feature is called wavefront healing \citep{Liang+etal2013} and it is produced by the decrease of the amplitude of the scattered wave relative to the incident wave with distance due to finite wavelength effects. These effects are observable when the ratio of sunspot radius to the wavelength of the incident wave is close to unity. As this ratio is reduced the scattered wave becomes more circular \citep[\eg][]{Zhao+etal2011}, and its energy is distributed in a wavefront whose size increases with the distance and, thus, its amplitude decreases. We find that in the spaghetti sunspot, the distance $x$ at which the magnitude of the travel-time shift starts to decrease is higher (Figure \ref{fig:traveltime_map_p1}). It may indicate that the radius that effectively produces the scattering of the monolithic sunspot is smaller than that of the spaghetti model. The same conclusion can be extracted from the examination of the curvature of the scattered wave (Figure \ref{fig:scattering_p1}). This agrees with the differences in the distribution of the magnetic flux between both models, since in the monolithic sunspot most of the magnetic flux is concentrated in the inner region while in the spaghetti model it is uniformly distributed (Figure \ref{fig:velocities_B_models}). Note that the $f$ mode travel-time shift (Figures \ref{fig:traveltime_map_f} and \ref{fig:traveltime_plot_f}) does not show significant wavefront healing inside the computational domain. The filter applied for those plots selects shorter wavelengths, reducing finite wavelength effects.

\citet{Schunker+etal2013} estimated that noise level in the time-distance measurements for a single sunspot for seven days is of order a few seconds (3.8 seconds for the $f$ mode and 3.5 seconds for the $p_1$ mode) after averaging over a region behind the sunspot around 60 Mm by 40 Mm. In our simulations, we find differences between models of order seconds when averaged over the bigger spatial region behind the sunspot allowed by the size of our computational domain. The noise level in the measurements must be a concern when detecting these changes in the travel-time shift.

\section{Fourier-Hankel analysis}
\label{sect:hankel}

The simulations were also analyzed by means of Fourier-Hankel analysis \citep{Braun+etal1988,Braun1995}. The vertical velocity wavefield is decomposed into radially ingoing and radially outgoing waves in an annular region. From the power and phase of those waves, we can measure the absorption $\alpha_{\rm m}(L)$ and phase shift ${\delta}_{\rm m} (L)$ produced by the magnetic structures as a function of the azimuthal order $m$ and the degree $L$. See \citet{Felipe+etal2012a} for details of the computations. We have evaluated the coefficients $A_m(L,\nu )$ and $B_m(L,\nu )$ for the complex amplitudes of the ingoing and outgoing waves, respectively, within an annular domain bounded by the radial distances $R_{\rm min}=12$ Mm and $R_{\rm max}=27$ Mm. This annular domain provides a resolution in spherical harmonic degree $\Delta L=291.54$. The duration $T$ of the simulation is given by the time that the wave packet needs to travel through the computational domain in the $x$ direction. In the case of the $f$ mode simulations $T=180$ min, which produces a sampling in frequency of $\Delta \nu=1/T=0.0959$ mHz. Since the $p_1$ modes travel faster, for those simulations $T=135$ min and the resolution in frequency is $\Delta \nu=0.1235$ mHz.

\subsection{Absorption coefficient}
\label{sect:absorption}

\begin{figure}[!ht] 
 \centering
 \includegraphics[width=9cm]{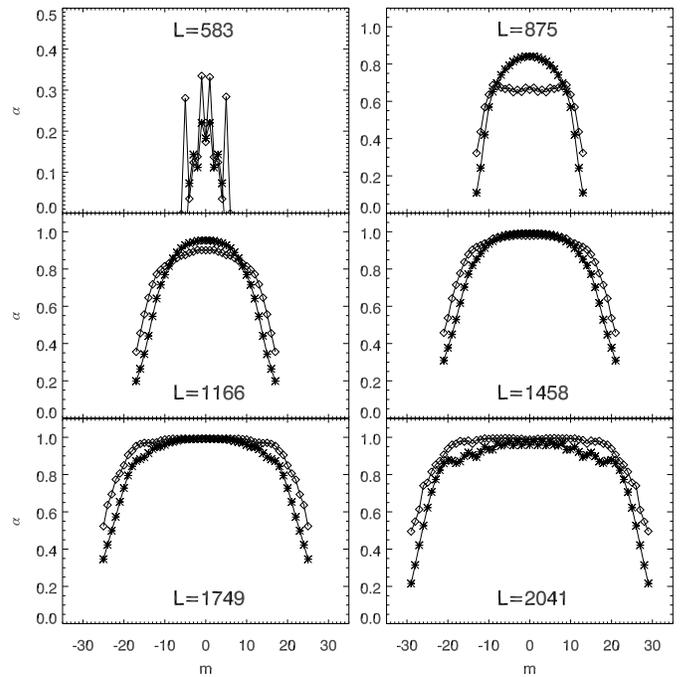}
  \caption{Absorption coefficient of the $f$ mode as a function of $m$ produced by the monolithic sunspot (asterisks) and the spaghetti sunspot (diamonds) at different degrees. Note the different scale of the top left panel.}
  \label{fig:absorption_f_l}
\end{figure}

The absorption coefficient $\alpha_m^n(L)$ is obtained as  

\begin{equation}
\label{eq:alpha}
\alpha_m^n(L)=1-|B_m^n(L)|^2/|A_m^n(L)|^2,
\end{equation}

\noindent where $|A_m^n(L)|^2$ and $|B_m^n(L)|^2$ are the power of the ingoing and outgoing waves, respectively, integrated along the frequencies that form the ridge of the mode with radial order $n$ ($n=0$ for the $f$ mode and $n=1$ for the $p_1$ mode) at degree $L$. Figure \ref{fig:absorption_f_l} shows a comparison between the absorption measured in the $f$ mode for the monolithic and spaghetti sunspots. The variation of $\alpha_m^0(L)$ with $L$ at different azimuthal orders is plotted. At a first glance, both sunspot models seem to produce a similar absorption pattern. The variation with $m$ is symmetric, with the highest absorption obtained at $m=0$. As $L$ increases, the distribution with $m$ becomes broader and the absorption increases. When the absorption is close to unity it saturates, and the profiles show a flat variation with $m$ at lower azimuthal orders followed by a sudden reduction. Looking at the comparison between both sunspots in detail, some differences arise. At $L=875$, the spaghetti sunspot shows a constant absorption for $|m|<10$. This flat distribution constrasts with the higher absorption produced by the monolithic model at those azimuthal orders. Moreover, for all $L$ values the spaghetti sunspot produces higher absorption at higher azimuthal orders. The measurement of the absorption coefficient as a function of the azimuthal order provides some limits to the horizontal extent of the absorbing area. \citet{Braun+etal1988} defined the impact parameter of the incident wave component as $m/k$, where $k=L/R_{\odot}$ is the horizontal wavenumber and $R_{\odot}$ is the solar radius. The observed absorption fall to zero when the impact parameter exceeds the radius of the absorbing region. In the case of the spaghetti sunspot, the broader distribution of the absorption as function of azimuthal order indicates that the absorbing region of this model is bigger, in agreement with the conclusions extracted from the travel-time maps.

\begin{figure}[!ht] 
 \centering
 \includegraphics[width=9cm]{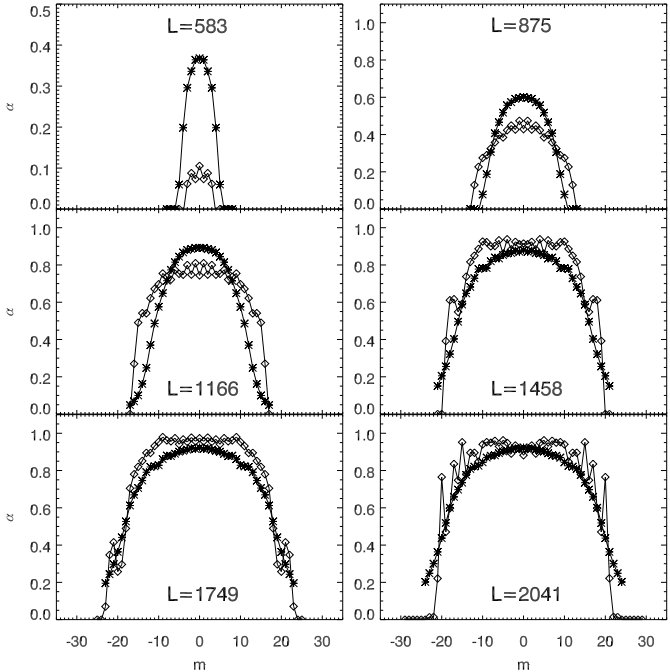}
  \caption{Absorption coefficient of the $p_1$ mode as a function of $m$ produced by the monolithic sunspot (asterisks) and the spaghetti sunspot (diamonds) at different degrees. Note the different scale of the top left panel.}
  \label{fig:absorption_p1_l}
\end{figure}

Figure \ref{fig:absorption_p1_l} shows the absorption coefficient measured in the $p_1$ mode for both sunspot models. In the same way as the $f$ mode measurements, the monolithic sunspot absorption is higher at low azimuthal orders, while in the spaghetti model the significant values of the absorption are extended to higher azimuthal orders. The difference between the absorption produced by both models at $L=583$ is striking and may demonstrate a way to distinguish the two models. In general, the absorption coefficient of the $f$ mode exceeds that of the $p_1$ mode, except the monolithic sunspot at $L=583$.

\subsection{Phase shift}
\label{sect:phase}

The phase shift ${\delta}_m^n(L)$ is given by

\begin{equation}
{\delta}_m^n(L) = \arg \left( \int_{\nu_n(L) - \delta \nu}^{\nu_n(L) + \delta \nu} 
 B_m^n (L, \nu) A^{n*}_m (L, \nu) d\nu \right),
\label{eq:phase}
\end{equation}

\noindent where $\nu_n(L)$ is the frequency of the mode $n$ at degree $L$, $\delta \nu$ indicates the range in frequencies around the ridge used for the integration, and the asterisk corresponds to the complex conjugate. The phase shift of the quiet Sun simulation is substracted from the phase shift of the sunspot simulations. According to this definition, a positive phase shift means that the phase of the wave is advanced in time by the scatterer. It indicates that the wave speed has been increased and, thus, corresponds to a negative travel time. The sunspot models analyzed in this work produce strong phase shifts, which are higher than 360$^o$ for some $L$ values. In Equation \ref{eq:phase}, the phase is obtained from the argument of a complex number through the arctangent, that is restricted to the interval [-180$^o$, 180$^o$]. For values out of this range the absolute phase cannot be retrieved and a phase wrap is produced. The signal was automatically unwrapped by detecting the phase jumps between two adjacent azimuthal orders at fixed $L$ higher (lower) than 180$^o$ (-180$^o$) and substracting (adding) 360$^o$.

\begin{figure}[!ht] 
 \centering
 \includegraphics[width=9cm]{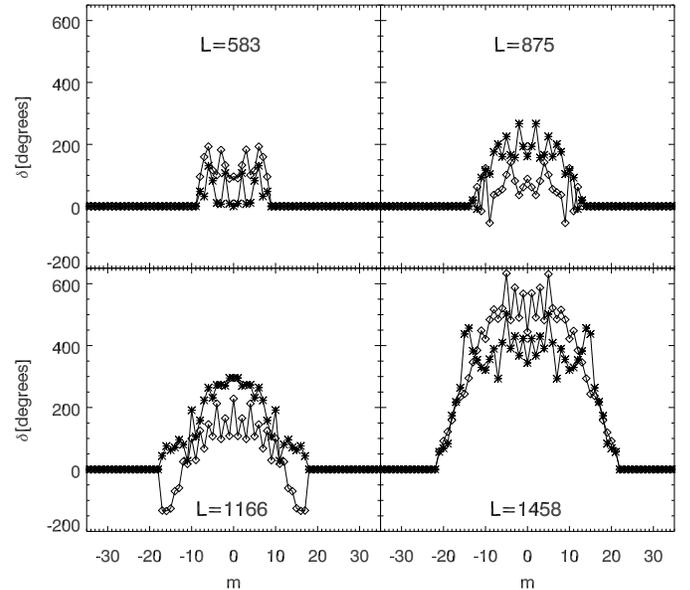}
  \caption{Phase shift of the $f$ mode as a function of $m$ produced by the monolithic sunspot (asterisks) and the spaghetti sunspot (diamonds) at different degrees.}
  \label{fig:phase_f_l}
\end{figure}

\begin{figure}[!ht] 
 \centering
 \includegraphics[width=9cm]{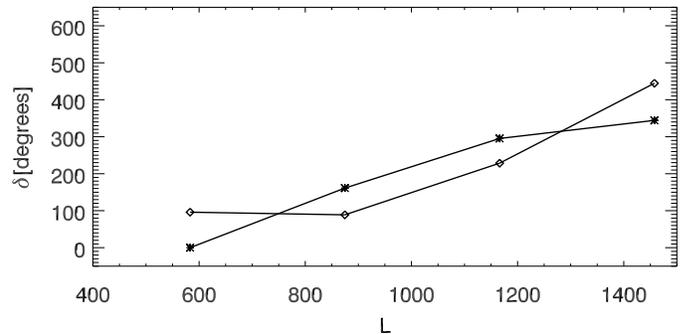}
  \caption{Phase shift of the $f$ mode as a function of $L$ produced by the monolithic sunspot (asterisks) and the spaghetti sunspot (diamonds) at $m=0$.}
  \label{fig:phase_f_m}
\end{figure}

Figure \ref{fig:phase_f_l} shows the variation of the phase shift with the azimuthal order at several $L$ values. The presented results are restricted to $L$ below 1500 because at higher $L$ values the multiple wrappings of the phase shift hinder the estimation of the absolute phase. In the case of the monolithic sunspot, the phase shift smoothly increases with $L$, and the distribution with azimuthal order becomes broader. Note that at $L=1458$, low azimuthal orders show a phase shift higher than one period. On the other hand, the phase shift of the spaghetti sunspot at low azimuthal orders shows small variations at the lower $L$ values, but it greatly increases at $L=1458$, being even higher than that produced by the monolithic model. These differences in the variation of the phase shift with $L$ are illustrated in Figure \ref{fig:phase_f_m}, where $\delta$ is plotted as a function of $L$ for $m=0$.

\begin{figure}[!ht] 
 \centering
 \includegraphics[width=9cm]{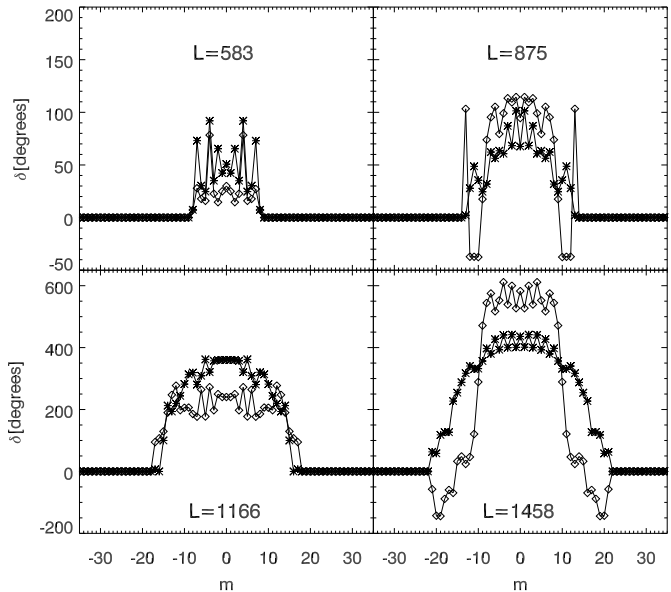}
  \caption{Phase shift of the $p_1$ mode as a function of $m$ produced by the monolithic sunspot (asterisks) and the spaghetti sunspot (diamonds) at different degrees.}
  \label{fig:phase_p1_l}
\end{figure}

\begin{figure}[!ht] 
 \centering
 \includegraphics[width=9cm]{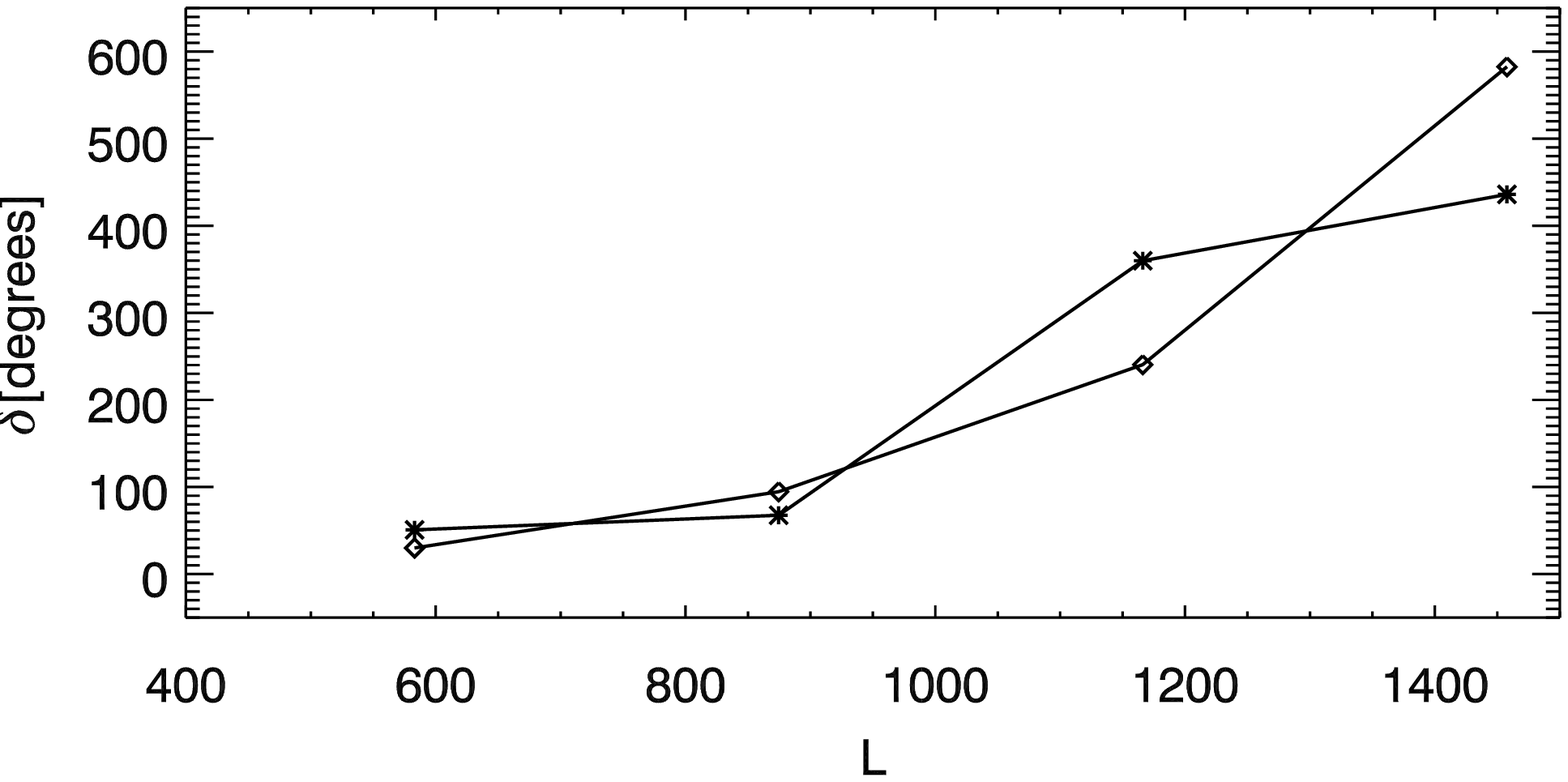}
  \caption{Phase shift of the $p_1$ mode as a function of $L$ produced by the monolithic sunspot (asterisks) and the spaghetti sunspot (diamonds) at $m=0$.}
  \label{fig:phase_p1_m}
\end{figure}

The phase shift of the $p_1$ mode shows a similar pattern (Figure \ref{fig:phase_p1_l}); increase with $L$ and the broadening in $m$. As shown in Figure \ref{fig:phase_p1_m}, for the $p_1$ mode the spaghetti sunspot also produces an steep enhacement of the phase shift between $L=1166$ and $L=1458$ in the lower azimuthal orders. At $L=1458$ and $|m|$ between 16 and 21, the spaghetti model shows some negative phase shifts, while in the case of the $f$ mode negative phase shifts appear at $L=1166$.

\section{Mode-mixing}
\label{sect:mode_mixing}

\begin{figure}[!ht] 
 \centering
 \includegraphics[width=9cm]{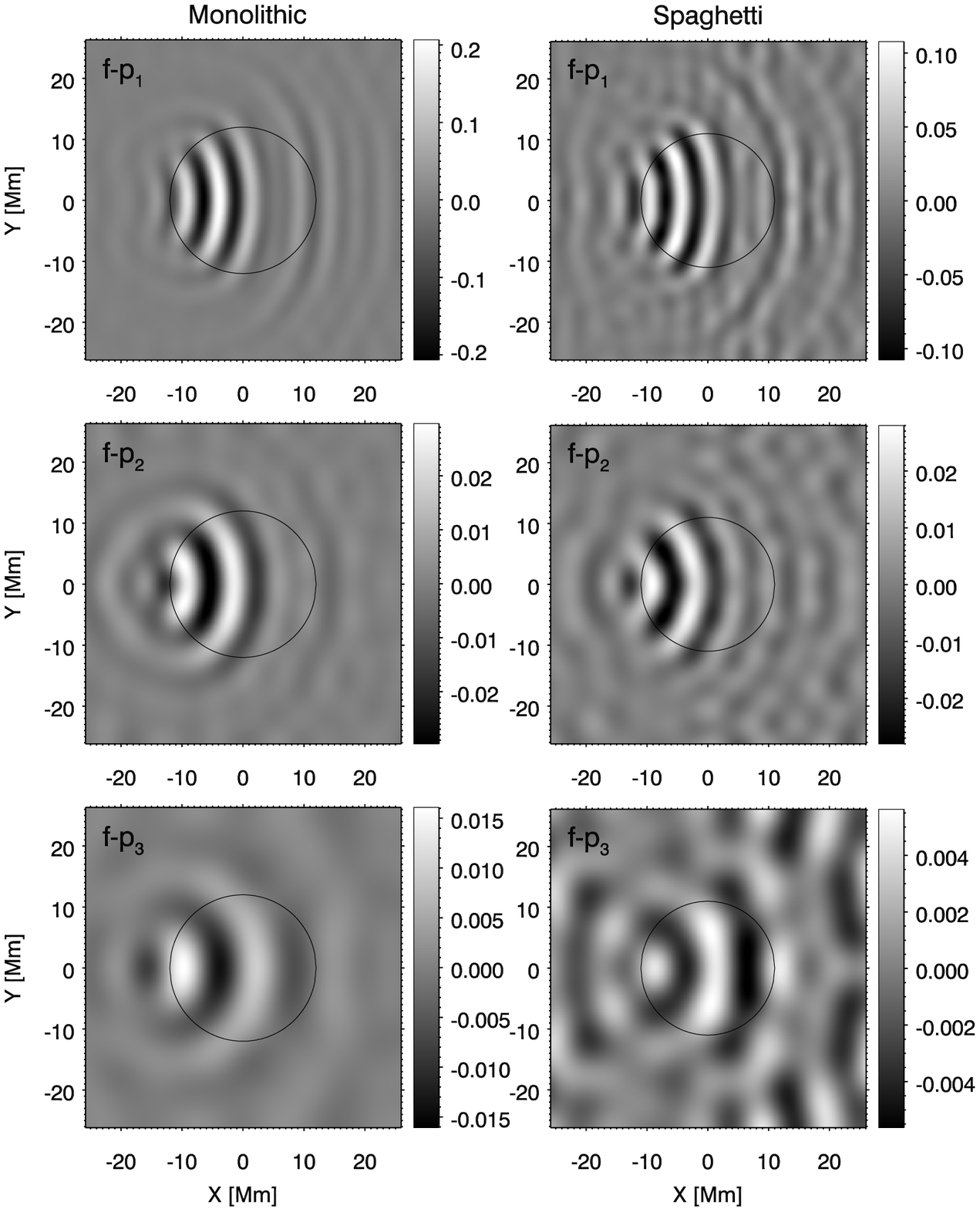}
  \caption{Spatial distribution of the scattered $p_1$ (top panels), $p_2$ (middle panels), and $p_3$ (bottom panels) modes produced by mode mixing from the incident $f$ mode after 70 minutes of simulation. Left column shows the monolithic model and right column the spaghetti model. The black circumferences indicates the limit of the sunspots. Note the differences in the gray scales. }
  \label{fig:mixing_f}
\end{figure}

\begin{figure}[!ht] 
 \centering
 \includegraphics[width=9cm]{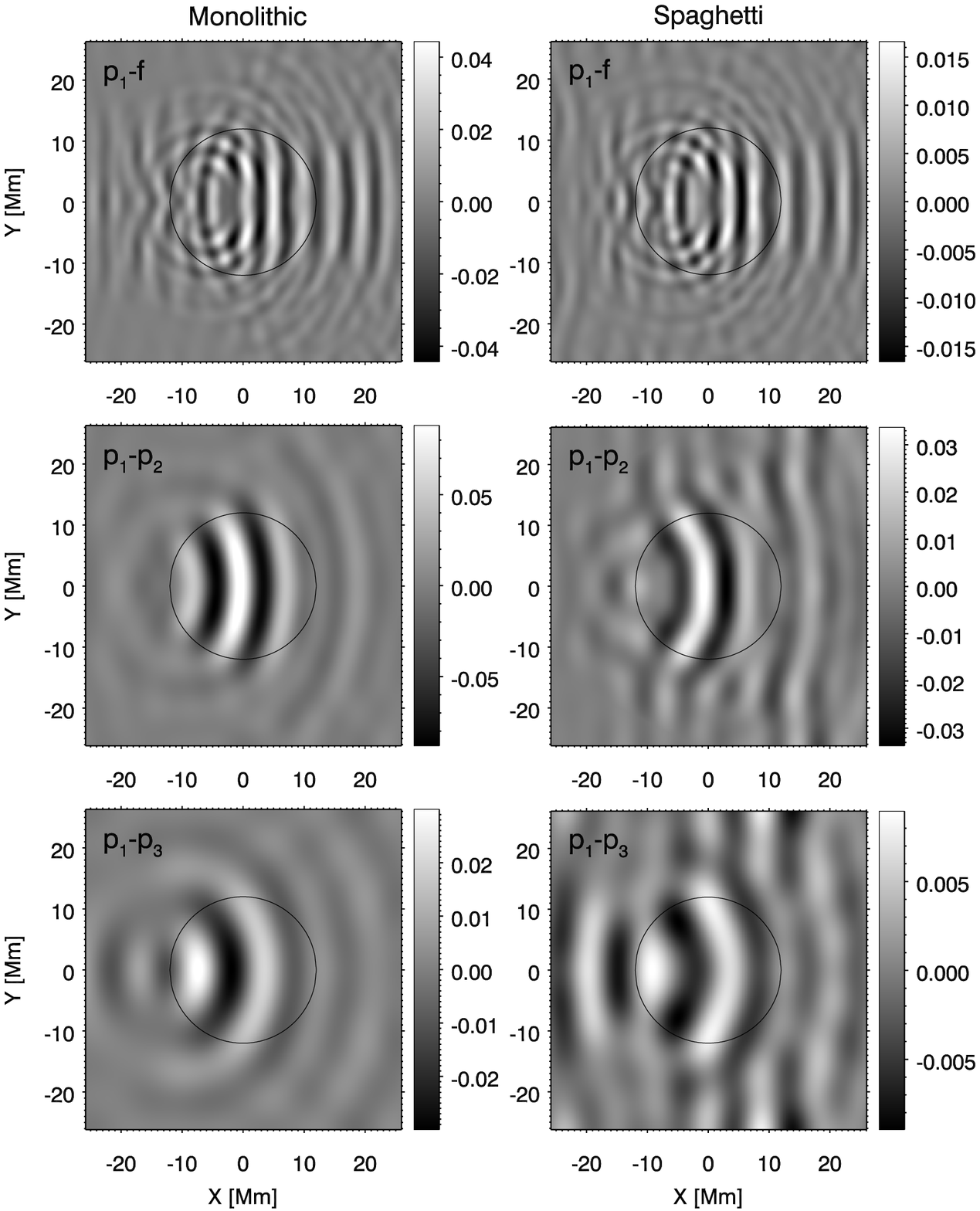}
  \caption{Spatial distribution of the scattered $f$ (top panels), $p_2$ (middle panels), and $p_3$ (bottom panels) modes produced by mode mixing from the incident $p_1$ mode after 50 minutes of simulation. Left columns show the monolithic model and right columns the spaghetti model. The black circumferences indicates the limit of the sunspots. Note the differences in the gray scales. }
  \label{fig:mixing_p1}
\end{figure}

For magnetic structures that are stationary in comparison to the typical wave period, the frequency of the scattered wave is the same as that of the incident wave. However, the interaction with the sunspot can convert an incident wave of radial order $n$ to a different outgoing radial order $n'$ through a process called ``mode-mixing'' \citep{D'Silva1994, Braun1995}. The outgoing wave keeps the frequency $\nu$ of the incident wave, but its wavenumber $k$ (or degree $L$) is changed from the corresponding for the ridge $n$ in the $k-\nu$ diagram to that of the ridge $n'$, according to the respective dispersion relations $\nu (n,k)$. 

Since in our simulations we introduce individual wave packets with fixed radial order as initial condition (either $f$ or $p_1$ modes), the power in the scattered waves in other radial orders must come from the mode-mixing from the radial order of the incident wave. We have filtered the vertical velocity wavefields in order to isolate the $f$, $p_1$, $p_2$, and $p_3$ modes in the frequency range between 3 and 4 mHz. Figures \ref{fig:mixing_f} and \ref{fig:mixing_p1} show maps of the filtered scattered waves produced by mode-mixing in the monolithic and spaghetti scenarios. The magnitude of the oscillations in these maps has been divided by the maximum amplitude of the filtered incident wave ($f$ and $p_1$ modes, respectively) in the quiet Sun reference simulation at the same time step in order to provide an approximation of what fraction of the incident wave is converted to a different radial order.

Figure \ref{fig:mixing_f} shows the mode-mixing produced from the $f$ mode ($n=0$) to higher order $p$ modes, from $p_1$ to $p_3$. There is a significant amount of mode-mixing from $f$ to $p_1$ mode, but its efficiency rapidly drops for $\Delta n=n'-n$ greater or equal to 2. The amplitude of the scattered $p_2$ is almost an order of magnitude lower than that of $p_1$, and for $p_3$ it is even smaller. A comparison of mode-mixing produced by the monolithic and spaghetti sunspots shows remarkable differences. This may be a way to distinguish monolithic and spaghetti models for sunspots. In the case of $f$ to $p_1$ scattering ($\Delta n=1$), the monolithic model generates a two times higher amplitude scattered wave. The $f$ to $p_2$ mode-mixing ($\Delta n=2$) is similar for both sunspots, but the amplitude of the $p_3$ mode ($\Delta n=3$) scattered by the spaghetti sunspot is also significantly lower, being barely distinguishable from the background.

With regards to the mode-mixing from an incident $p_1$ mode (Figure \ref{fig:mixing_p1}), the most efficient scattering corresponds to $\Delta n=1$, that is, to the scattering from $p_1$ to $p_2$. The amplitude of scattered waves produced by mode-mixing with $\Delta n=-1$ and $\Delta n=2$ is smaller. In a similar fashion to the $n=0$ case, the monolithic sunspot also produces a notable higher amount of mode-mixing in comparison to the spaghetti model. For all $\Delta n$ analyzed, the amplitude of the scattered wave produced by the monolithic model is around 3 times higher than that of the spaghetti sunspot.

\section{Discussion and conclusions}
\label{sect:conclusions}

In this paper, we have studied the scattering produced by the interaction of $f$ and $p_1$ modes with two different sunspot models: monolithic and spaghetti sunspots. Although our simple models do not account for the full complexity of solar magnetic structures, they retain the basic properties that characterize each kind of sunspot model. Our work is a first step to a qualitatively understanding of the differences in the seismic response between both models, rather than a quest for an agreement with observational measurements. We have evaluated the wave interaction with the model magnetic structures by means of the analysis of the wavefield as well as performing local helioseismic measurements, including travel-time shift maps and Fourier-Hankel analysis. The results reveal some differences in the seismic signature of those sunspot models: 
     
1. The wavefront of the waves scattered by the monolithic sunspot has more curvature, while the waves scattered by the spaghetti sunspot are more flat (Figures \ref{fig:scattering_f} and \ref{fig:scattering_p1}).

2. Behind the sunspot, the spaghetti sunspot produces approximately constant phase shift for all the $y$ positions where the flux tubes are present, while the magnitude of the travel time shift generated by the monolithic model is higher behind the center of the sunspot (bottom panel of Figure \ref{fig:traveltime_plot_f}).

3. The travel time shift produced by the monolithic sunspot peaks closer to the axis of the spot than the travel time shift produced by the spaghetti model (top panel of Figure \ref{fig:traveltime_plot_p1}).

4. The distribution of the absorption coefficient with azimuthal order is broader for the spaghetti sunspot (Figures \ref{fig:absorption_f_l} and \ref{fig:absorption_p1_l}).

5. The absorption of the $p_1$ mode produced by the monolithic sunspot at low $L$ is much higher than that of the spaghetti sunspot.

6. The efficiency of mode-mixing is strikingly higher for the monolithic sunspot (Figures \ref{fig:mixing_f} and \ref{fig:mixing_p1}).

7. The spaghetti model generates negative phase shifts at high azimuthal orders for some $L$ values (Figures \ref{fig:phase_f_l} and \ref{fig:phase_p1_l}).

8. The variation of the phase shift with $L$ produced by the spaghetti sunspot between $L=1166$ and $L=1458$ is steeper than that of the monolithic sunspot (Figures \ref{fig:phase_f_m} and \ref{fig:phase_p1_m}).

These differences suggest that helioseismology techniques could distinguish between the sunspot models analyzed in this work. However, our aim is to find a measurement that can detect above the noise level the properties that are inherent of the sunspot model type (monolithic or spaghetti). The differences listed from 1 to 4 suggest that the horizontal extent of the scattering region of the monolithic sunspot is smaller than that of the spaghetti model. Those results agree with the fact that in the monolithic sunspot the magnetic field is stronger at the axis of the sunspot and decreases with the radius, while in the case of the spaghetti model the flux tubes are uniformly distributed inside a 10.7 Mm radius (Figure \ref{fig:velocities_B_models}). This way, points 1 to 4 probably cannot be used to discern between monolithic and spaghetti sunspots. Nevertheless, the coherence among all these measurements and the characteristics of the sunspot models serves as a sanity check of the calculations.   

Regarding points number 5, previous studies of the scattering produced by bundles of flux tubes have pointed out that multiple scattering enhances the absorption \citep{Keppens+etal1994, Hanasoge+Cally2009, Felipe+etal2013}. Those works are restricted to a small number of flux tubes. Our simulations of the spaghetti sunspot model account for a realistic number of flux tubes for the first time. The stronger efficiency of the monolithic model at low $L$ values constrasts with the results suggested by earlier work. It must be noted that in this paper we are comparing different magnetic structures. In our monolithic sunspot the absorption is produced by the conversion of the incident acoustic waves into field guided magnetoacoustic modes (slow, fast, or Alfv\'en), whereas for the spaghetti sunspot the absorption is caused by the excitation of tube waves \citep[for further discussion see][]{Felipe+etal2012a}. The differences in the absorption between the two sunspot models at $L$ values below 1000 (Figures \ref{fig:absorption_f_l} and \ref{fig:absorption_p1_l}) can be detected above the observational noise level if the signal to noise is increased by averaging over azimuthal orders \citep{Braun+etal1988,Bogdan+etal1993,Braun1995}. 

The recent work by \citet{Couvidat2013a} shows the absorption coefficient for 15 sunspots and reveals substantial differences among them. This result suggests that the differences in the absorption coefficient between sunspots of the same kind (monolithic or spaghetti) may be higher than those obtained between the two models analyzed in this work. Furthermore, the individual properties of each sunspot, regarding size and magnetic field strength, may produce a stronger influence on the absorption than the nature of the subsurface magnetic structure. One way to proceed in the future consists of combining helioseismic measurements with direct polarimetric observations in order to set some constraints on the absorption expected from the observed magnetic flux.

As we listed in point number 6, the measured mode-mixing shows differences between both models. The absorption coefficient provides a measure of the power lost by an specific mode (with radial order $n$, frequency $\nu$, degree $L$). However, not all this power corresponds to a real absorption produced by magnetic field guided waves that extract energy from the acoustic cavity. If there is a significant amount of mode-mixing, part of the energy is also scattered to other wave modes with different radial order $n'$ and degree $L'$, but with the same frequency since the sunspot is stable. The evaluation of mode-mixing in both sunspot models indicates that a remarkable amount of power is scattered to other radial orders, particularly for $\Delta n=1$ and for an incident $f$ mode, and that the mode-mixing generated by the monolithic sunspot is up to 3 times higher than that of the spaghetti model. Recently, \citet{Zhao+Chou2013} have observationally obtained the scattered wavefunctions from $n$ to another radial order $n'$. The relation between this kind of measurements and the absorption coefficient might be a valuable criterion to discern between monolithic and spaghetti scenarios.

The phase shift measurements show some of the most noticeable differences between both models, listed as numbers 7 and 8. On the one hand, the spaghetti sunspot produces negative phase shifts at high $m$ values. For the $p_1$ mode, these negative phase shifts appear at $L=875$ and $L=1458$, while in the case of the $f$ mode they are obvious at $L=1166$ and slightly visible at $L=875$. On the other hand, the comparison of the magnitude of the phase shift between both models depends on the mode and $L$ value considered. At $L$ below 1200 the phase shift of the monolithic model is generally higher, except the $f$ mode at $L=583$ and the $p_1$ mode at $L=875$. However, at higher $L$ the phase shift produced by the spaghetti sunspot is significanly higher for both $f$ and $p_1$ modes. \citet{Felipe+etal2013} found that multiple scattering tends to reduce the phase shift, and according to \citet{Hanasoge+Cally2009} the extent of the region of influence of the near field is $\lambda/2$. Since the spaghetti sunspot shows a steep phase shift enhacement at higher $L$ values (lower wavelengths), our results point out that for the longer wavelengths the phase shift of the spaghetti sunspot is reduced due to multiple scattering effects, while at shorter wavelengths the reduction of the relevance of those effects generates a sudden increase in the phase shift. This way, the variation of the phase shift as a function of $L$ (wavelength) also provides a hint to distinguish between the models.

Note that the difference in the $L$ dependence of the phase shifts between the spaghetti and monolithic models is also affected by the depth dependence of the two models. Higher $L$ waves are sensitive to shallower regions of the atmosphere. As shown in Figure \ref{fig:velocities_B_models}, around the photosphere the sound speed of the monolithic model is reduced, while our spaghetti sunspot does not account for a realistic Wilson depression. The work by \citet{Lindsey+etal2010} has evaluated the effects of the cooling of the umbral subphotosphere on the travel-time shifts, finding that the Wilson depression produces a reduction of the travel-time, while \citet{Schunker+etal2013} have assessed the sensitivity of travel-times to the Wilson depression. The development of magnetohydrostatic spaghetti sunspot models with a realistic Wilson depression is a challenging problem. One possible way to proceed in the future is to construct such models by extracting their properties from magnetohydrodynamic numerical simulations of sunspots \citep[\eg][]{Rempel+etal2009,Rempel2011}, and experimenting with bottom boundary conditions that produce spaghetti-like configurations for the subsurface magnetic field.

The differences in the depth dependence and magnetic field inclination also affect mode-mixing \citep{D'Silva1994}. While the magnetic field of the monolithic sunspot spreads with height, with its inclination increasing with the radial distance, the fluxtubes from the spaghetti model are almost vertical. This difference may contribute to the lower efficience of mode-mixing in the spaghetti sunspot. The depth dependence and the inclination of the magnetic field are not necessarily a consequence of the type of sunspot model, and calls attention to be cautious with using the variation of the phase shift with $L$ and the mode-mixing measurements to discern between monolithic and spaghetti sunspot scenarios.
    
Some limitations hinder our ability to make meaningful comparisons of the presented results and actual solar observations. Several published studies of the absorption and phase shift produced by sunspots are restricted to $L$ lower than 600 \citep{Braun+etal1992, Braun1995}, while our analysis focuses on higher $L$ values. In order to extend our results to those low $L$ values, it is necessary to perform the simulations using bigger horizontal computational domains and increase the annular region used for the Hankel analysis. Measurements at higher $L$ values show a decrease in the absorption for $L\gtrsim 500$ \citep{Braun+etal1988, Bogdan+etal1993, Couvidat2013a}. Observations are affected by background convective (or instrumental) noise and finite lifetime effects, but we made no attempt to include those features in the numerical calculations. Future simulations should also study higher order $p$ modes and examine more detailed sunspot models, especially in the case of the spaghetti sunspot. We have used the same vertical flux tube all over the sunspot region. Improved models should account for the radial variation of the magnetic flux inside the sunspot, the Wilson depression, and also include a variety of flux tubes with different radius, magnetic field strengths, and inclinations.

\acknowledgements  This research has been funded by NASA
 through projects NNH09CE43C, NNH09CF68C, NNH09CE41C, NNH07CD25C, and NNX14AD42G. Computer resources were provided by the Extreme Science and Engineering Discovery Environment (XSEDE), which is supported by National Science Foundation grant number OCI-1053575, and NASA's Pleiades supercomputer at Ames Research Center. This work uses Hankel decomposition software developed by D. C. Braun (NWRA) and publicly available at http://www.cora.nwra.com/$\sim$dbraun/hankel/. ACB acknowledges DFG SFB 963 ``Astrophysical Flow Instabilities and Turbulence'' (project A18) aimed at understanding solar and stellar dynamos.

\end{document}